\documentclass[showpacs]{revtex4}
\usepackage{amsmath}
\usepackage{latexsym}
\usepackage{epsfig}

\begin{document}

\title{Migration and proliferation dichotomy in tumor cell invasion}
\author{Sergei Fedotov${}^1$ and Alexander Iomin${}^2$ }
\affiliation{${}^1$ Department of Mathematics, University of Manchester,
Manchester M60 1QD, UK \\
${}^2$ Department of Physics, Technion, Haifa, 32000, Israel}

\begin{abstract}
We propose a two-component reaction-transport model for the
migration-proliferation dichotomy in the spreading of tumor cells. By
using a continuous time random walk (CTRW) we formulate a system of the
balance equations for the cancer cells of two phenotypes with random
switching between cell proliferation and migration. The transport process is
formulated in terms of the CTRW with an arbitrary waiting time distribution
law. Proliferation is modeled by a standard logistic growth. We apply
hyperbolic scaling and Hamilton-Jacobi formalism to determine the overall
rate of tumor cell invasion. In particular, we take into account both normal
diffusion and anomalous transport (subdiffusion) in order to show that the
standard diffusion approximation for migration leads to overestimation of
the overall cancer spreading rate.
\end{abstract}

\pacs{05.40.-a, 05.40.Fb, 87.15.Vv, 87.17.Ee, 82.39.Rt}
\maketitle

Extensive investigations have been devoted to the modeling of cancerous
growth (see, for example, reviews \cite{preziosi1,swanson,ECMTB05}, and
references therein). Although a great deal of progress has been made in this
theory, especially for solid tumors, for which growth is basically due to
cell proliferation, our understanding of malignant gliomas, the diffusive
and highly invasive brain tumors, is much less complete (see, a review \cite%
{swanson}). The main reason for this is that unlike solid tumors,
gliomas not only are able to proliferate but also to invade the
surrounding brain parenchyma actively. The surgical resection of
diffusive tumors is
ineffective since the cancer cells have already invaded the surrounding
brain tissue. This leads to recurrence of tumor, and the prognosis for
patients suffering from malignant gliomas is very poor. Thus
proliferation and especially migration of gliomas provide a significant
challenge for modelling, and this is why the invasiveness of tumors
has been studied extensively in recent years (see, for example,
\cite{swanson,Giese1,glioma}).

Invasion, itself, is a very complex process of receptor-mediated
transport \cite{karsten}, which involves several steps of cell
migration and proliferation (see a review \cite{Giese1}). Experimental
evidence indicates the lower proliferation rate of migratory cells in
comparison with the tumor core, which indicates an inverse correlation
between mobility and
proliferation of cell population. The existence of this important
phenomenon was supported by
numerous experimental data obtained \emph{in vitro} and clinical data
obtained \emph{in vivo} \cite{Giese1}. It was formulated by Giese et al.
\cite{Giese2} as a \textit{migration-proliferation dichotomy}. It turns out
that proliferation and migration of tumor cells are mutually exclusive
phenotypes: the spreading suppresses cell proliferation and visa versa. The
molecular mechanism for this dichotomy has been suggested in \cite{wells};
and then an active implementation for the numerical modelling of the
brain tumor and its fractional topology has been established
\cite{mansury}. It turns out that this behavior of cells is an
inherent process of a so-called continuous time random walk (CTRW). This
transport concept, based on jump and waiting time distributions, has
been extensively and
successfully employed for numerous applications \cite{Mon,klafter}.
Migration-proliferation dichotomy was formulated in the framework of the
CTRW in \cite{iom1}. The primary focus was on the influence of cell
fission on transport properties of cells. An essential decrease in cell
motility during fission time, or their self-entrapping, is determined by
the interaction of cells with their environment.
\emph{In vitro} experimental observations of cell transport confirm
the essential decrease in cell motility during cell proliferation
\cite{Dieterich}.

Usually the random mobility of tumor cells is described by Fick's
law. However, it has been shown that the diffusion approximation for the
transport process together with a logistic growth yields an overestimation
of the overall propagation rate \cite{F1,FM}. Since the tumor cells'
migration is the most critical feature of brain cancer, causing treatment
failure, the transport has to be properly understood. Therefore we need to
extend the diffusion analysis by introducing a more realistic description of
the transport of mobile tumor cells. It is one of the main purposes of this
Letter to take into account anomalous transport (subdiffusion) leading to
slow mobility of cancer cells in the invasive zone.

In this Letter we propose an alternative approach for the
migration-proliferation dichotomy. We employ a two-component CTRW,
assuming
that the glioma cells are of two phenotypes. In state $1$ (migratory
phenotype) the cells randomly move but there is no cell fission. In state $2$
(proliferating phenotype) the cancer cells do not migrate and only
proliferation takes place. The exact mechanism of switching between the two
phenotypes is not known. An interesting deterministic mechanism for this
phenotype switch has been suggested recently in \cite{athale}. However,
the mathematical modeling involves many parameters,
some of which are difficult to estimate. Here we propose the stochastic
approach for the proliferation-migration switching that involves only two
parameters. We assume that the cell of type $1$ remains in a state $1$
during a waiting time $\tau _{1}$ and then switches to a cell of type
$2$. After a waiting time $\tau _{2}$, spent in a state $2$, it switches
back to a cell of type $ 1 $. Both waiting times $\tau _{1}$ and
$\tau _{2}$ are mutually independent random variables. In this paper we
consider them exponentially distributed with parameters $\beta _{1}$ and
$\beta _{2}$: namely, $f(\tau _{i})=
\beta _{i}\exp \left( -\beta _{i}\tau _{i}\right) $. \ Here the
parameters $\beta_{i}$ are the switching rates, namely, $\beta _{1}$ is
the switching rate from state $1$ to $2$, while $\beta _{2}$ determines
the transition rate $2\rightarrow 1$. The last parameter can control the
lower migratory cell proliferation. Indeed, if the average time
$\langle\tau_2\rangle=1/\beta_2$
is much less than time of cell doubling then the proliferating cells in
state $2$ do not have enough time to proliferate even with the same high
rate as the core of cells.

Let us introduce the density for the cells of migratory phenotype (cells of
type $1$), $n_{1}(t,x)$, and for the cells of proliferating phenotype (cells
of type $2$), $n_{2}(t,x)$. The balance equations can be written as follows
\begin{eqnarray}
n_{1}(t,x) &=&n_{1}(0,x)\Psi (t)e^{-\beta _{1}t}+\int_{0}^{t}\int
n_{1}(t-s,x-z)\rho (z)\psi (s)e^{-\beta _{1}s}dzds  \notag \\
&&+\beta _{2}\int_{0}^{t}n_{2}(t-s,x)\Psi (s)e^{-\beta _{1}s}ds\,,
\label{n1}
\end{eqnarray}%
\begin{eqnarray}
n_{2}(t,x) &=&n_{2}(0,x)e^{-\beta
_{2}t}+U\int_{0}^{t}n_{2}(t-s,x)(1-n_{2}(t-s,x)/K)e^{-\beta _{2}s}ds  \notag
\label{n2} \\
&&+\beta _{1}\int_{0}^{t}n_{1}(t-s,x)e^{-\beta _{2}s}ds\,,
\end{eqnarray}%
where $\rho (z)$ is the probability density for migration jump length, while
$\psi (s)$ is the probability density of waiting times between jumps, and
$ \Psi (t)=1-\int_{0}^{t}\psi (s)ds$ is the probability that a cell of
type $1$ makes no jump until time $t$. The exponential factor
$e^{-\beta_{i}t}=\int_{t}^{\infty }f(\tau _{i})d\tau _{i}$ is the
probability that
cells of phenotype $i$ do not switch until time $t$. Eq. (\ref{n1})
describes the balance of cells of type $1$ at time $t$ at position $x$. The
first term on the right hand side of the equation
represents those cells of type $1$\ that stay up to time $t$ at position $x$
such that no jump occurs, and no switch $\ 1\rightarrow 2$ takes place.
The
independence of the random jumps and switching gives us the probability
$ \Psi (t)e^{-\beta _{1}t}$ while the first factor $n_{1}(0,x)$ is the initial
density of cells of type $1$.
The second term
describes the number of cells of type $1$ arriving at $x$ up to time $t$ due
to the following random mechanism of migration: the cell of type $1$ at time
$t-s$ at position $x-z$ waits a random time $s$ before jumping at a distance $z$
and remains a cell of type $1$. This process is determined by the transition
probability $\psi (s)\rho (z)$. The limits of the space integral are
determined by the boundaries. The last term in Eq. (\ref{n1})
represents the number of cells of type $2$ that switches to the cell of type
$1$ up to time $t$ and remaining the cells of type $1$ (due to the factor $
e^{-\beta _{1}s}$). It also takes into account the fact that if transition $
2\rightarrow 1$ happens at time $t-s$, then no jump takes place during the
remaining time $s$ (due to the factor $\Psi (s)$).

Regarding Eq. (\ref{n2}), the first term on the right hand side has
the same physical meaning as one in Eq. (\ref{n1}). The second term
is the logistic growth \cite{Murray} for cells of type $2$, which occurs
providing that no switch takes place up to time $t$. Here $U$ is the cell
proliferation rate and $K$ is the carrying capacity of the environment. The
last term of Eq. (\ref{n2})
represents the number of cells of type $1$ switching to the state $2$ over
the time interval $(0,t)$. Note that one-component balance equation
involving transport and production term has been analyzed in \cite{FM}.

The balance equations (\ref{n1}) and (\ref{n2}) can be written as the system
of integro-differential equations. By using the Laplace transform
$\tilde{n}_{i}(H)=\int_{0}^{\infty}e^{-Ht}n_{i}(t)dt$ and presenting the
left hand side of the
equations in the form $H\tilde{n}_{i}(H)-\tilde{n}_{i}(0)$ which is the
Laplace transform of the time derivative, one obtains
\begin{equation}
\frac{\partial n_{1}}{\partial t}=\int_{0}^{t}\alpha (t-s)\int \left[
n_{1}(s,x-z)-n_{1}(s,x)\right] \rho (z)dzds-\beta _{1}n_{1}+\beta
_{2}n_{2}\,,  \label{intdef}
\end{equation}%
\begin{equation}
\frac{\partial n_{2}}{\partial t}=Un_{2}(1-n_{2}/K)+\beta _{1}n_{1}-\beta
_{2}n_{2}\,,  \label{dn2}
\end{equation}%
where the `memory' kernel $\alpha (t)$ is defined in terms of its Laplace
transform
\begin{equation}
\tilde{\alpha}(H)=\frac{\left( H+\beta _{1}\right) \tilde{\psi}(H+\beta _{1})%
}{\left( 1-\tilde{\psi}(H+\beta _{1})\right) }\,,\text{ \ \ }
\label{LaplaceMemory}
\end{equation}%
with $\tilde{\psi}(H)=\int_{0}^{\infty }\psi (t)e^{-Ht}dt$. Note that the
equivalence of one-component balance equation to a master equation
involving
memory kernel has been shown in \cite{Mon}. It should be emphasized that it
is impossible to find an explicit expression for memory kernel $\alpha (t)$
for arbitrary choices of waiting-time pdf $\psi (t)$. In what follows we
will be concerned with the overall rate of the spreading of gliomas. It
turns out that this rate depends on the Laplace transform $\tilde{\alpha}(H)$
rather than $\alpha (t)$. That is why the formula (\ref{LaplaceMemory})
plays a crucial role in this Letter.

It is natural to assume that jumps of migrating cells are small and there is
no convection ($\int z\rho (z)dz=0$). Expanding $n_{1}(s,x-z)$ in the Taylor
series in Eq. (\ref{intdef}), we obtain
\begin{equation}
\frac{\partial n_{1}}{\partial t}=\frac{\sigma ^{2}}{2}\int_{0}^{t}\alpha
(t-s)\frac{\partial ^{2}n_{1}}{\partial x^{2}}ds-\beta _{1}n_{1}+\beta
_{2}n_{2}\,,  \label{memory}
\end{equation}%
where $\sigma ^{2}=\int z^{2}\rho (z)dz$. Generalization on $3D$ is
straightforward, namely, the second derivative is replaced by the Laplace
operator $\Delta $.

Now we are in a position to find the overall rate at which the cancer cells
spread. The main purpose here is to find the dependence of the rate of
invasion on the statistical characteristics of the random switching process,
$\beta _{1}$ and $\beta _{2},$ and random walk in space, $\sigma ^{2}$ and $%
\psi (t)$. We expect that the system of equations together with appropriate
initial conditions has a traveling wave solution (planar front) with some
velocity $u$ common to both densities $n_{1}$ and $n_{2}$. The objective
here is to find the rate $u$ without resolving the shape of the traveling
waves \cite{F1,Fr}. For this purpose we use a hyperbolic scaling $%
x\rightarrow x/\varepsilon ,\,t\rightarrow t/\varepsilon $ and the rescaled
densities $n_{i}^{\varepsilon }\left( t,x\right) =n_{i}\left( t/\varepsilon
,x/\varepsilon \right) $. We apply the exponential transformation
\begin{equation}
n_{i}^{\varepsilon }\left( t,x\right) =A_{i}\exp \left( -\frac{G\left(
t,x\right) }{\varepsilon }\right) \,,\quad i=1,2\,,  \label{wkb}
\end{equation}
where positive constant $A_{1}$ and $A_{2}$ represent the stable equilibrium
points of the densities $n_{1}^{\varepsilon }$ and $n_{2}^{\varepsilon }$.
Our purpose is to find an equation for $G\left( t,x\right) $ which gives us
the spreading front position $x\left( t\right) $ in the limit of the
long-time and large-distance, from the equation $G\left( t,x\left( t\right)
\right) =0$ \cite{F1}. To ensure the minimal spreading rate we use the
front-like initial conditions: $n_{i}(0,x)=A_{i}$ for $x<0,$ and
$n_{i}(0,x)=0$ for $x\geq 0$ \cite{Fr}. Substituting (\ref{wkb}) into the
equations for the densities $n_{1}^{\varepsilon }$ and
$n_{2}^{\varepsilon }$, one obtains two equations for $A_{1}$ and $A_{2}$
in the limit $\varepsilon \rightarrow 0.$ This system has a non-trivial
solution when the corresponding determinant is equal to zero. This yields
a generalized Hamilton-Jacobi equation, involving two first
derivatives $\partial
G/\partial t$ and $\partial G/\partial x:$
\begin{eqnarray}
&&\left[ 1-\left( 1+\frac{\sigma ^{2}}{2}\left( \frac{\partial G}{\partial x}%
\right) ^{2}\right) \int_{0}^{\infty }e^{\frac{\partial G}{\partial t}s}\psi
(s)e^{-\beta _{1}s}ds\right] \left[ 1-U\int_{0}^{\infty }e^{\frac{\partial G%
}{\partial t}s}e^{-\beta _{2}s}ds\right]  \notag \\
&&-\beta _{1}\beta _{2}\int_{0}^{\infty }e^{\frac{\partial G}{\partial t}%
s}\Psi (s)e^{-\beta _{1}s}ds\times \int_{0}^{\infty }e^{\frac{\partial G}{%
\partial t}s}e^{-\beta _{2}s}ds=0.  \label{final}
\end{eqnarray}%
Note that inferring Eq. (\ref{final}), we do not make any assumptions
regarding waiting time pdf $\psi (t)$. If we introduce the Hamiltonian
function $H=-\partial G/\partial t,$ the generalized momentum $p=\partial
G/\partial x$, and the Laplace transform $\tilde{\psi}(H)=\int_{0}^{\infty
}\psi (t)e^{-Ht}dt$, then the Hamilton-Jacobi equation (\ref{final}) takes
the form
\begin{equation}
\frac{\sigma ^{2}p^{2}}{2}=\frac{1}{\tilde{\psi}(H+\beta _{1})}\left[ 1-%
\frac{\beta _{1}\beta _{2}(1-\tilde{\psi}(H+\beta _{1}))}{(H+\beta
_{1})\left( H+\beta _{2}-U\right) }\right] -1.  \label{mom}
\end{equation}%
The latter equation is important, since it allows us to find the
overall spreading rate $u=\min_{H}\left\{ H/p(H)\right\} $ by using \cite{F1}
\begin{equation}
u=\frac{H}{p(H)},\;\;\;\frac{\partial p}{\partial H}=\frac{p(H)}{H}.
\label{HE}
\end{equation}
In the symmetrical $3D$ case, Eq. (\ref{mom}) corresponds to the Hamiltonian
motion in the radial direction. Let us illustrate the use of the above
theory through two typical distributions for the waiting-time pdf
$\psi (t)$.

First, we consider a probability distribution function for the
\textit{exponentially distributed waiting times:}
$\psi (t)=\tau^{-1}e^{-t/\tau }$. We find $\tilde{\psi}(H)=(1+H\tau)^{-1}$
and $\tilde{\alpha}(H)=\tau ^{-1}$, and therefore
$\alpha (t)=\tau ^{-1}\delta (t)$. This corresponds to the
classical Fick's law for transport with the diffusion coefficient
$D=\sigma ^{2}/2\tau $. Thus we have a classical system of reaction-diffusion
equations such that the equation for the migratory cells is
\begin{equation}
\frac{\partial n_{1}}{\partial t}=D\frac{\partial ^{2}n_{1}}{\partial x^{2}}%
-\beta _{1}n_{1}+\beta _{2}n_{2}\,.  \label{dif1}
\end{equation}%
\ \ The momentum $p(H)$ can be found from (\ref{mom})%
\begin{equation}
p^{2}=\frac{(H+\beta _{1})}{D}-\frac{\beta _{1}\beta _{2}}{D\left( H+\beta
_{2}-U\right) }.  \label{momen}
\end{equation}%
If we assume that $\beta _{1}=\beta _{2}$, we can find from (\ref{HE})
and (\ref{momen}) $p=\left( U/D\right) ^{1/2}$, and $H=U$. Therefore, the
spreading rate is $u_{0}=\left( UD\right) ^{1/2}$ which is half of the
classical Fisher-KPP propagation speed. This is a very interesting result
showing that the propagation rate is independent of the random
migration-proliferation switching when cell transport is the Brownian motion
and $\beta _{1}=\beta _{2}$. When $\beta _{1}\neq \beta _{2}$ one can find
the ratio of the propagation rate $u$ and $u_{0}=\left( UD\right) ^{1/2}$ as%
\begin{equation}
\left( \frac{u}{u_{0}}\right) ^{2}=\frac{H^{2}\left( H+\beta _{2}-U\right) }{%
U\left( \left( H+\beta _{2}-U\right) (H+\beta _{1})-\beta _{1}\beta
_{2}\right) }.  \label{nor}
\end{equation}

The situation changes for the \textit{\ power law distribution
(anomalous transport):} $\psi (t)\sim (\tau /t)^{1+\gamma }$ with $0<\gamma
<1$. This is the case when the mean waiting time is divergent: $<t>=\infty .$
This assumption alone leads to the temporal fractional differential operator
and corresponding anomalous diffusion equation \cite{klafter}. The mean
squared displacement for mobile cells is
\begin{equation}
<x^{2}(t)>=\frac{4D_{\gamma }}{\Gamma (1+\gamma )}t^{\gamma },\text{ }
\label{an}
\end{equation}%
where $D_{\gamma }=\sigma ^{2}/2\tau ^{\gamma }$ is the generalized
diffusion coefficient with the dimension $cm^{2}s^{-\gamma }.$ One of the
main aims of this Letter is to find the overall propagation of cancer cells
as a result of interaction of the anomalous migration (\ref{an}), logistic
proliferation and random migration-proliferation switching. For this purpose
it is more convenient to define $\psi (t)$ by its Laplace transform $\tilde{%
\psi}(H)=\left( 1+\left( H\tau \right) ^{\gamma }\right) ^{-1}$ \cite%
{klafter}, such that the momentum $p(H)$ can be found from (\ref{mom})
\begin{equation}
p^{2}=\frac{(H+\beta _{1})^{\gamma }}{D_{\gamma }}-\frac{\beta _{1}\beta
_{2}(H+\beta _{1})^{\gamma -1}}{D_{\gamma }\left( H+\beta _{2}-U\right) }.
\label{fr}
\end{equation}%
This formula together with (\ref{HE}) allows us to find the overall
propagation
rate of tumor cells $u_{\gamma }$ in the fractional diffusion case. It is
clear that the case $\gamma =1$ corresponds to the normal diffusion
approximation for cell migration (see (\ref{momen})). One can find from
(\ref{HE}), (\ref{momen}) and (\ref{fr}) the ratio of the anomalous propagation
rate $u_{\gamma }$ and the normal rate $u$ determined by (\ref{nor}):
\begin{equation}
\frac{u_{\gamma }}{u}=(H_{\gamma }\tau +\beta _{1}\tau )^{\frac{1-\gamma }{2}%
}\, ,  \label{last}
\end{equation}
where $H_{\gamma }$ is the solution of  $\partial p/\partial H=p(H)/H$.
Since the ``microscopic'' time $\tau $ is much smaller than the
``mesoscopic''
reaction time $U^{-1}$ and switching time $\beta _{1}^{-1}$ and $H_{\gamma
}\sim U$, we conclude that $H\tau +\beta _{1}\tau <1.$ It follows from
(\ref{last}) that the ratio $u_{\gamma }/u$ increases with $\gamma $ in the
interval $0<\gamma <1$. This means that the standard diffusion
approximation
leads to overestimation of the overall cancer spreading. It is clear from
these two examples of normal and anomalous diffusions that the advantage of
balance Eqs. (\ref{n1}) and (\ref{n2}) is that they are related to
``mesoscopic'' description of migratory cancer cells, and give us the
statistical meaning of the reaction-diffusion equations or fractional
equations that are introduced usually phenomenologically

In summary, we present a two-component model for \textit{a
migration-proliferation dichotomy} in the spreading of tumor cells in the
invasive zone. We use a probabilistic approach based on the CTRW theory
for migration, logistic growth and random \textit{proliferation-migration}
switching with exponentially distributed waiting times. Our approach is not
restricted to the specific mechanism of proliferation described by a
logistic growth. Moreover, Eq. (\ref{n2}) for proliferation can be
accompanied by a nutrient control or chemotaxis \cite{khain}. The main point
of the paper is that cancer cell transport is subdiffusive rather than
diffusive described by Fick's law (the cancer cells are not Brownian
particles!). The advantage of our approach is that it allows us to take into
account anomalous (subdiffusive) transport within the general scheme of
migration, proliferation and phenotype switching. We show the equivalence of
balance equations to a system of master equations involving memory kernels
for the transport of mobile cells. By using a hyperbolic scaling and
Hamilton-Jacobi formalism we derive formulae for the overall spreading rate
of cancer cells. We show that the memory effects (subdiffusion) leads to a
decrease in propagation rate compared to a standard diffusion
approximation
for transport. An analytical expression for the memory kernel can be
obtained for more complicated processes. For example, for the family of
gamma distributions with parameters $m$ and $\tau $, $\psi (t)=\tau
^{-m}t^{m-1}e^{-t/\tau }/\Gamma (m)$ \cite{FO}. We have $\tilde{\psi}%
(H)=\left( 1+H\tau \right) ^{-m}$ and
\begin{equation*}
\tilde{\alpha}(H)=\frac{\left( H+\beta _{1}\right) }{(1+H\tau +\beta
_{1}\tau )^{m}-1}\,.
\end{equation*}%
For example, if $m=2$, then $\tilde{\alpha}(H)=\tau ^{-1}\left( 2+H\tau
+\beta _{1}\tau \right) ^{-1}$ and the memory kernel is $\alpha (t)=\tau
^{-2}e^{-(2+\beta _{1}\tau )t/\tau }.$ The integro-differential Eq.
(\ref{memory}) can be written as the hyperbolic reaction-transport equation,
and corresponding traveling wave solutions can be found in \cite{MFo}.
Renovation processes with arbitrary probability densities for switching
waiting times will be considered in the future publications.

\renewcommand{\theequation}{A-\arabic{equation}} \setcounter{equation}{0}

\end{document}